\documentstyle[12pt]{article}
\advance\leftskip by -20pt \advance\rightskip by -25pt
\advance\vsize by 4.0 truecm
\advance\voffset by -1.0 truecm
\begin{document}
\baselineskip14pt
\begin{titlepage}
\thispagestyle{empty}
\date{}
\title{Isotropization of Bianchi-Type Cosmological Solutions
 in Brans-Dicke Theory }
\author{P. Chauvet\thanks{e-mail:pcha@xanum.uam.mx} \\
 Departamento de F\'{\i}sica, \\
Universidad Aut\'onoma Metropolitana--Iztapalapa \\
P. O. Box. 55--534, M\'exico D. F.\\
C. P. 09340 MEXICO \\
Fax 52-5-724 4611
\and
J.L.Cervantes-Cota\thanks{e-mail:
jorge@spock.physik.uni-konstanz.de}  \\
 University of Konstanz, p.o.box: 5560, M 678 \,\ \\
D-78434 Konstanz \\
Fax 49-7531-88 3888}
% Short title: Isotropization in Brans-Dicke Theory
\maketitle
PACS number: 04.20.Jb, \hfill gr-qc/9502015
%\end{titlepage}
%
%%%%%%%%%%%%%%%%%%%%%
%%\section*{Abstract}
%%%%%%%%%%%%%%%%%%%%%
\setcounter{page}{1}
\begin{abstract}
\baselineskip12pt
The cosmic, general analitic solutions of the Brans--Dicke Theory for the
flat space of homogeneous and isotropic models containing perfect, barotropic,
fluids are seen to belong to a wider class of solutions --which includes
cosmological models with the open and the closed spaces of the
Friedmann--Robertson--Walker metric, as well as solutions for
models with homogeneous but anisotropic spaces corresponding to the
Bianchi--Type metric clasification-- when all these solutions are expressed
in terms of reduced variables. The existence of such a class lies in the fact
that the scalar field, $\phi$, times a function of the mean scale factor or
``volume element'',  $a^3 = a_1  a_2 a_3$, which depends on time and on the
barotropic index of the equation of state used, can be written as a function
of a ``cosmic time'' reduced in terms of another function of the mean scale
factor depending itself again on the barotropic index  but independent of the
metrics here employed.
This reduction procedure permites one to analyze if explicitly given
anisotropic cosmological solutions ``isotropize'' in the course of their time
evolution. For if so can happen, it could be claimed that there exists a
subclass of solutions that is stable under anisotropic perturbations.
\end{abstract}
\end{titlepage}
\baselineskip17pt
%%%%%%%%%%%%%%%%%%%%%%%%%%%%%%%%%%%%%%%%%
%\newpage
\setcounter{page}{2}
\section{INTRODUCTION}
The first authors to realize the possibility of giving a reason for the
viability of the ``cosmological principle'' without the necessity of imposing
 highly special initial conditions before the ``inflationary programme'' was
developed, were Hoyle and Narlikar \cite{HoNa63}.
However, their explanation came also before the cosmic
microwave background (CMB) radiation
was discovered, and nowadays, most investigators believe that the steady state
theory is untenable from the observations.
Other investigators within Einstein's general relativity theory (EGR), notably
Misner \cite{Mi67}, tried to demonstrate --unsuccesfully-- that the large
scale structure of the Universe, in particular its isotropy, could be
attributed to the nature of the matter processes, such as dissipation,
that took place at a very early stage of development of the Universe
independent of its initial conditions (chaotic cosmology), that is,
that the Universe lost memory of any
initially imposed anisotropy or inhomogeneity (Barrow and Matzner
\cite{BaMa77},
Doroshkevich {\sl et. al.\/} \cite{DoZeNo68},
Misner \cite{Mi68}, Rees \cite{Re72}, Zel'dovich and Starobinsky
\cite{ZeSt72}).
More to the point, within theoretical cosmology in general
and in the context of EGR in particular, one is still looking for a
satisfactory explanation to the following observational facts:
The formation of galaxies and cluster of galaxies which means that the
universe is not homogeneous locally, and on the other hand the CMB
radiation that seems to be very nearly isotropic on account of its
Planck spectrum and the lack of structure in its intensity, from which it has
been concluded that the large--scale structure of the actual Universe must be
homogeneous and isotropic. The standard big bang model, thought to give the
most accurate description of the Universe, has the peculiarity that it
appears to need a very special set of initial conditions to be viable. This
state of affairs have produced several studies in different but related
directions to obtain reasonable explanations to this conundrum, like the
``inflationary programme'' which nowadays is a popular approach, nevertheless
not without some drawbacks, to solve also some other problems in cosmology
(for a present review see Olive \cite{Ol90}).
A most important reason why cosmological models that predict inflation
in the early universe are interesting is the hope they will
explain the observed state of the universe without appeal to highly
special initial conditions. Even so, most inflationary cosmological models
have assumed Friedmann--Robertson--Walker (FRW) symmetry from the outset. The
horizon size in the FRW models suggests the possibility that physical
interactions could have homogenized and isotropized the universe, and
therefore that its present state could have evolved from more general
initial conditions. Yet, it is not established if general cosmological models
with non FRW initial conditions that actually enter, and gracefully exit, an
inflationary phase initial homogeneity and anisotropy will be smoothed out
since gravitational interactions tend to enhance inhomogeneities instead of
smoothing them out, see Wald \cite{Wa83}.

On experimental grounds the degree of anisotropy of our Universe is
somewhat bounded by the current measurements of
the CMB anisotropy made by COBE \cite{sm92}. Nevertheless, present
observations are still far from elucidating
all the properties that an actual model of the universe should have. Even the
degree of anisotropy of the primeval radiation or the counts of
radio sourcess or galaxies in the various directions in the sky involve some
uncertainties and, agreement between the observed chemical composition and the
predictions of the Friedmann models merely signifies that the time rate of
change of the volume occupied by matter and the rate of expansion of a
Friedmann model concide. On the other hand, astronomical observations seem to
imply that isotropization must occur at a rather early epoch, maybe even for
$z>>10^9$. It is feasible that for particular anisotropic models this could be
achieved for the instantaneous values of the anisotropy parameters. But, the
case is that there are observational properties that depend on the degree of
anisotropy over an extended period of time, again like the CMB isotropy.

In EGR, without the aid of a cosmological constant or inflation, Collins
and Hawking \cite{CoHa73} examined the question in terms of
an ``initial conditions'' analysis. They obtained that the set of spatially
homogeneous cosmological models approaching isotropy in
the limit of infinite times is of zero
measure in the space of all spatially homogeneous models, which in turn implies
that the isotropy of the models is unstable to homogeneous and anisotropic
perturbations. Yet, other authors define the concept of isotropization
in a different way (see Novikov \cite{No72}, Mac Callum \cite{Ma79} and
Zel'dovich and Novikov \cite{ZeNo83}).
Therefore, in the literature concerned with the mathematical analysis of
anisotropic models the term isotropization is often mentioned but its
precise definition is author dependent.
For the Bianchi--Type models it is claimed that a positive  cosmological
constant provides an effective mean of isotropizing homogeneous universes
(Wald \cite{Wa83}). However, in this context Barrow \cite{Ba87} has shown that
contrary to previous expectations, perfect fluid cosmologies need not approach
isotropy and homogeneity as $t \to \infty$.

Following the general line of thought on this subject put forward by
Zel'dovich and Novikov, one declares that homogeneous cosmological models that
isotropize are those that ``approach to a Friedmann model in the course of time
as the universe expands'', which means that ``its geometric and dynamic
parameters as well as those concerning the distribution and motion of matter
and radiation are nearly the corresponding quantities in a Friedmann model''.
Accordingly, one assumes that the idea conveyed by isotropization is the
property that an arbitrary solution possesses to model our Universe which
permits it to evolve from an initial, general state, into a state that is
presently isotropic on a ``large scale'' and is, therefore, well described by
a Friedmann solution.  We reduce the general scope of the problem by assuming
initial homogeneity, limiting ourselves to test the isotropization properties
of certain specific non inflationary solutions in the Jordan-Brans-Dicke (JBD)
cosmological theory. So, if at its outset the universe was not in
isotropic expansion, the above ideas imply that one can examine, on a first
approximation, the properties of homogeneous but anisotropic models assumed to
describe correctly the early stages of its expansion. Of these, only those
Bianchi--Type models whose group type comprise FRW models may isotropize:
Types I, V, VII$_0$, VII$_h$ and IX.

In this paper the above concept of isotropization is dealt with in a direct,
but admittedly limited, way by qualifying and quantifying it through a
"Raychaudhuri type" equation common to all Bianchi--Type models: Given an
explicit solution, one can directly check if it may or may not approach to a
Friedmann regime in the course of its cosmological time evolution,
specifically, if the different anisotropic scale factors of a Bianchi model
in the various directions approach arbitrarily near to a unique, single
function of time. By this procedure one can then answer the question, at
least for some representative spatially homogeneous models of the Bianchi--
Type (I, V, IX), of whether, and if so, how in the JBD cosmological theory
a present large scale isotropy resulted from an initially anisotropic but
homogeneous expanding universe.

\section{FRW FIELD EQUATIONS}
The JBD field equations for the FRW cosmology with a barotropic, perfect fluid,
$p \,=\, \beta \rho \,$, $\, -1 < \beta < 1$~ (the ~$ \beta = {1\over 3}$,
equation of state for incoherent radiation or ultrarelativistic matter is
excluded) are
\begin{equation}
\label{rho}
\rho a^{3(1+\beta)} = M_\beta, \hskip 2cm  M_\beta = {\rm const. }
\end{equation}
\begin{equation}
\label{a1}
3 (1-\beta){a^\prime \over a}  \,=\, \left( {\psi^{\prime} \over \psi} \right )
 - {{ (1 - 3 \beta)m_\beta \eta
+ \eta_0} \over \psi}~~,\quad m_\beta = {{8 \pi M_\beta}
\over {3 + 2 \omega}}~~.
\end{equation}
The dynamic equation is
\begin{equation}
\label{d1}
 \left( {\psi^{\prime \prime} \over \psi} \right) - {{[2(2-3\beta)+
3 (1-\beta)^2\omega]}m_\beta\over \psi} =
{{-6 (1-\beta) k} \over {a^{2(1 - 3 \beta)}}}~~, \quad k=0,~\pm 1
\end{equation}
and the constraint equation is
{\renewcommand{\arraystretch}{3}\begin{equation}\hspace*{-1.25cm}
\begin{array}{c}
\label{con1} \displaystyle
{3 \over {2(1-\beta)} } \left( {\psi^{\prime \prime} \over \psi}
\right)-{1\over{(1-\beta)^2}}\left({\psi^{\prime} \over \psi} \right )^2 -
{{(1-3 \beta)}\over{(1-\beta)^2}} \left( {{ (1-3 \beta)m_\beta\eta + \eta_0}
\over \psi} \right) \left( {\psi^{\prime} \over \psi} \right )  \\
\displaystyle
+{{[2-3 \beta+ {3\over2}(1-\beta)^2\omega]}
\over{(1-\beta)^2}} \left({{ (1-3 \beta)m_\beta\eta + \eta_0}
\over \psi} \right)^2
+ {{3[2+\omega(1-\beta)(1+3 \beta)]m_\beta}\over {2(1-\beta) \psi}}
  \\ \displaystyle   = 0 \,\ , \hfill \displaystyle
\end{array}\end{equation}}
where $\psi~ \equiv~ \phi a^{3 (1 - \beta)}$, $\phi$ is the JBD scalar field,
$a$ is the scale factor, $\omega$ the coupling parameter of the theory,
 $\eta$ the ``cosmic time parameter", $\eta_0$ an integration constant, and
 $(\,)^\prime = \partial_{\eta}$, where $dt \,=\, a^{3 \beta} d \eta $~
 (for details see Chauvet and Pimentel \cite{ChPi92}, and references therein).

For $k = 0$ equation (\ref{d1}) is directly integrated. One gets
\begin{equation}
\label{psi}
\psi = A \eta^2 + B \eta + C \end{equation}

where A, B and C  are constants  such that
\begin{equation}
\label{consta}
A=\left[2-3\beta+{3 \over 2}(1-\beta)^2 \omega\right] m_\beta~~~.
\end{equation}
Substitution of (\ref{psi}) and (\ref{consta}) in the constraint
equation hands out the
following results. The constant $B$ is undetermined and so, up to $B$, $C$
also remains undetermined. Therefore, three different possible cosmic
solutions to the FRW flat space  $(k=0)$ exist distinguished by the sign of
the determinant, $\Delta~ \equiv~ B^2 - 4AC$, which itself depends on the
relation between the equation of state, through $\beta$, and the coupling
parameter $\omega$ in a rather complex way. The behavior of the scalar
field $\phi$ implies, for each type of determinant, $\Delta > 0$, $\Delta < 0$
and $\Delta = 0$, the possible existence of two branches: essentially, ones
with $\phi$ an increasing function of time and the others with $\phi$ a
decreasing function of time (the solutions are given explicitly and
thoroughly discussed by Gurevich et.al. \cite{GuFiRu73}, Ruban and
Finkelstein \cite{RuFi75} and Morganstern \cite{Mo71}).

$\phi$ is obtained by the straightforward integration of
\begin{equation}
\label{phi}
 {\phi^\prime \over \phi} = {{(1-3\beta)m_\beta\eta + \eta_0} \over
{\psi}}~~, \end{equation}
and the scale factor, $a$ , found from it through the definition of
 $\psi$:
\begin{equation}
\label{a3}
 {a^{3(1-\beta)}} = {{A\eta^2+B\eta+C} \over {\phi}}~~~.
\end{equation}

Equation (\ref{con1}) does not involve the curvature constant $k$
explicitly and so,
$\psi = A\eta^2 + B\eta + C$ is also a solution to actually both
the open and closed space dynamic equation (\ref{d1}) provided that

\begin{equation}
\label{phia}
 \phi a^{(1+3\beta)} ={2 + (1-\beta)(1+3\beta)\omega \over 2 (1+3\beta)k }
m_{\beta}  \,\ .  \end{equation}
The same as in the flat space case, $A$, $B$ and  $C$ are obtained from
the constriction equation (\ref{con1}).  For both, $k = +1$ and $-1$,
it is valid that
\begin{eqnarray}
\label{abc}
A &=& {- {(1-3\beta)^2m_\beta} \over {(1+3\beta)}} ~~,
 \nonumber\\
B &= &- 2\left({  {1-3\beta}\over{1+3\beta} }\right) \eta_0~~,
\\
C &=& - {\eta^2_0 \over (1+3\beta) m_{\beta}}~~~~~.
 \nonumber
\end{eqnarray}
Its determinant is then
\begin{equation}
\label{dis}
   \Delta =  B^2 - 4AC  = 0  . \end{equation}

The explicit solutions for these two models are
\begin{equation}
\label{phisol}
 \phi = { -1 \over {(1 + 3 \beta)m_\beta}} \left({[2+(1-\beta)(1+3\beta)
\omega] {m_\beta}^2} \over {-2k} \right)^{3(1-\beta)\over {2(1-3\beta)}}~~
[(1-3\beta)m_\beta \eta + \eta_0]^{-{{1+3\beta} \over{ 1-3\beta}}}~~,
\end{equation}
and
\begin{equation}
\label{asol}
 a =  \left( { {-2k} \over [2+(1-\beta)(1+3\beta)\omega]{m_{\beta}^2}}\right)
^{1 \over 2(1-3\beta)} [(1-3\beta)m_\beta \eta+\eta_0]^{1 \over {1-3\beta}}~~.
\end{equation}

The solutions for the JBD flat space were previously obtained
by several authors (Gurevich {\sl et. al.\/} \cite{GuFiRu73},
Morganstern \cite{Mo71} and references therein, and in another context
by Chauvet and Pimentel \cite{ChPi92}).

Next, we present the anisotropic Bianchi field equations in the above variables
in order to analyze later their asymptotic solutions.
\section{ANISOTROPIC FIELD EQUATIONS}
Three extra equations, and simple modifications to the FRW equations~
(\ref{rho})--(\ref{con1}) presented above, describe the Bianchi--Types I, V
and IX examined in this paper. Equations (\ref{rho}) and (\ref{a1}) remain
formally the same, while equation (\ref{d1}) gets its ``curvature'' term
modified
and is then written
\begin{equation}
\label{d1m}
 \left( {\psi^{\prime \prime} \over \psi} \right) - {{ [2 (2 - 3 \beta) +
3(1-\beta)^2\omega ]m_\beta} \over
\psi} \,=\, (1-\beta) a^{6 \beta} \,\, {}^* R_j ~~. \end{equation}
The constriction equation is a ''Raychaudhuri type" equation so that the left
 hand side of equation (\ref{con1}) remains unaltered, but instead of
being equal to zero as in the FRW cosmology, it is in this case:
{\renewcommand{\arraystretch}{3}\begin{equation}\hspace*{-1.25cm}
\begin{array}{c}
\label{con2}\displaystyle
{3 \over {2(1-\beta)}}\left( {\psi^{\prime \prime} \over \psi} \right)-
{1\over{(1-\beta)^2}}\left( {\psi^{\prime} \over \psi} \right )^2 -
{{(1-3 \beta)}\over{(1-\beta)^2}} \left( {{ (1-3 \beta)m_\beta\eta + \eta_0}
\over \psi} \right) \left( {\psi^{\prime} \over \psi} \right ) \\
\displaystyle
+{{[2-3 \beta+ {3\over2}\omega(1-\beta)^2]}
\over {(1-\beta)^2}} \left({{ (1-3 \beta)m_\beta\eta + \eta_0}
\over \psi}\right)^2+{{3 [2+\omega(1-\beta)(1+3\beta)] m_\beta}
\over{2(1-\beta) \psi}} \\  \displaystyle
 =- (H_1 - H_2)^2 - (H_2 - H_3)^2 - (H_3 - H_1)^2 \equiv ~
\sigma ( \eta) \,\ .
\end{array}\end{equation}}

$\sigma$ is, for short, the ``shear''. $\sigma = 0$, is a
necessary condition to obtain a FRW cosmology since it implies $H_1 = H_2 =
H_3$ ( see Chauvet et.al. \cite{ChCeNu91}). If the sum of the squared
differences of the Hubble expansion rates tends to zero it would mean that
anisotropic scale factors tend to a single function of time which is,
presumably, the scale factor of a corresponding Friedmann model.  However, in
general not all Bianchi models contain a FRW space-time.

The three extra equations describe the dynamical evolution of the
``anisotro\-pic scale factors'' $a_1, a_2$ and $a_3$~:
\begin{equation}
\label{a123}
 (\psi H_i)^\prime \,=\, [1 + (1 - \beta) \omega] m_\beta + \psi \,\,
a^{6 \beta}{}^*R_{ij} ~~. \hskip .5cm {\rm i=1, 2, 3}
\end{equation}
{}From equations (\ref{d1m}) through (\ref{a123}), and
for the rest of this paper, we use
the following notation and conventions: $a$ is presently the mean scale factor,
$a^3 = a_1 a_2 a_3$, the $H_i$'s $i = 1, 2, 3$ are the Hubble expansion
rates, $H_i = a_i^\prime / a_i$, ${}^*R_j$ is the ``spatial three--curvature''
that belongs to a given Bianchi--Type model, ${}^*R_j = \sum
\limits_{i=1}^3~{}^*R_{ij}$ is a column sum, and the ${}^*R_{ij}$ are
``partial curvature''~
terms pertaining to specific scale factor dynamic equations, in our case:
\begin{equation}
\label{curv}
\matrix {
& &~{}^I &~~~~~~~~~{}^V~~~~~~~~ &~{}^{IX} \cr
& & & & \cr
& &0 & 2  / a_1^2 &
[a_1^4-a_2^4-a_3^4 + 2 a_2^2 a_3^2] / (-2 a^6) \cr
& & & & \cr
& {}^*R_{ij}~ = &0 & 2  / a_1^2 & [a_2^4-a_3^4-a_1^4 + 2 a_1^2 a_3^2]
/ (-2 a^6) \cr
& & & & \cr
& &0 & 2  / a_1^2 & [a_3^4-a_1^4-a_2^4 + 2 a_1^2 a_2^2] / (-2 a^6) \cr}
\end{equation}

In the next section it is first shown that $ \psi = A \eta^2 + B \eta + C $
includes solutions for the above, purposely chosen homogeneous but anisotropic,
cosmological models.  It will be shown that the obtained $ \psi $ class
of solutions consist of two parts: the isotropic and the anisotropic one. Then,
it will be clear that the latter approach to zero as the cosmic time parameter
evolves, i.e., these solutions tend asymptotically to their corresponding
isotropic group solutions, the FRW models.

\section{ANISOTROPIC SOLUTIONS AND THEIR
\vskip 0.1cm \hskip 0.7em ASYMPTOTIC BEHAVIOR}

We show next that $ \psi \,= \,A \eta^2 + B \eta + C $ is a solution for the,
homogeneous but anisotropic, Bianchi--Type cosmological models.
\bigskip

\noindent {\bf Bianchi- type I}
\par
For this Bianchi type model, ${}^*R_I=0$. It is direct to see by substituting
\begin{eqnarray}
\psi = A \eta^2 + B \eta + C  \nonumber &&
\end{eqnarray}
into equation (\ref{d1m}), that $A$ has the same expresion as the
one given by equation (\ref{consta}) (from now on we attach a subindex
to the $A$, $B$ and $C$ to distinguish between the different
Bianchi models):
\begin{equation}
\label{consta1}
 A_{{}_I} \,=\, [2-3 \beta+{3\over 2}(1-\beta)^2\omega]m_\beta~~~~.
\end{equation}

 By  direct substitution of the above results into equation (\ref{con2})
one finds
that $B_{{}_I}$ remains undetermined and may be put equal to any convenient,
but arbitrary value, and that
\begin{eqnarray}
\label{restabc}
\lefteqn{3 (1 - \beta )^2(3 + 2 \omega ) m_\beta \, C_{{}_I} = }
\nonumber \\ &&
-  { ( 1 - \beta )^2 \over \omega ^3} (h_1^2 + h_2^2 + h_3^2) -
m_\beta \eta ^2_0 A_{{}_I} +{B_{{}_I}}^2 + (1 - 3 \beta ) m_\beta \eta_0
B_{{}_I} ~~,
\end{eqnarray}
where the $h_i$'s are constants such that
\begin{equation}
\label{hi}
 H_i= {1 \over 3} {{a^3}^{\prime} \over a^3} + {h_i \over \psi } ~~,
\end{equation}
The non-vanishing constants $h_i$'s determine the anisotropic character of
the solutions. They obey the condition
\begin{equation}
\label{resth123}
 h_1 + h_2 + h_3 = 0  . \end{equation}
\par

By integration of equation (\ref{hi}), using equation (\ref{a1})
and equation (\ref{psi}), one
finds explicitly $a = a(\eta)$ (first obtained by Ruban and
Finkelstein \cite{RuFi75}, see also Chauvet and Guzm\'an \cite{ChGu86}
 and Chauvet \cite{Ch83}).

For the Bianchi's  in general, when equation (\ref{psi}) is substituted into
the
Raychaudhuri equation (\ref{con2}), one obtains the shear
as a function of $\psi$ and
the $h_i$'s,
\begin{equation}
\label{sig}
 \sigma(\eta) = - {3(h_1^2 + h_2^2 + h_3^2) \over 2\psi ^2} .
\end{equation}

This term permits, beside the only one allowed $\Delta =0$ solution for
the $k \not=0$, FRW models, two other solutions with $\Delta \not= 0$ such that
$\sigma \to 0$ as $\eta \to \infty$  (or $t \to \infty$).  It is in this sense
that these solutions may isotropize in the course of their time evolution (note
a mathematical characteristic of the anisotropic solutions shown by
the above results, i.e., the relation between the exponents $h_{i's}$ and
the $B$ and $C$ coeficients).

\vskip .4cm
\noindent {\bf Bianchi- type V}
\par
Equation (\ref{psi}) is a solution for this Bianchi model, with $A$, $B$ and
$C$
equal to
\begin{eqnarray}
\label{constab5}
A_{{}_V} & \,=\, & - {{(1-3\beta)^2m_\beta}\over{(1+3\beta)}}~ \\
B_{{}_V} & \,=\, & -2 \left({{1-3\beta}\over{1+3\beta}}\right) \eta_0 ~.
\nonumber
\end{eqnarray}
$A_{{}_V}$ and $B_{{}_V}$ are equal to the ones obtained for the isotropic,
 $k=\pm 1$ cases, but
\begin{equation}
\label{rest5}
 m_\beta \, (1+3\beta) \, C_{{}_V} \,=\, -\,
{(1+3\beta)^2 (h_1^2 + h_2^2 + h_3^2) \over 18 \beta + \omega (1+3\beta)^2}
\,-\, \eta_0^2~~. \end{equation}
So that,
\begin{equation}
\label{phi5}
 \phi= \left[{[2+(1-\beta)(1+3\beta)\omega]m_\beta}\over{-2(1+3\beta)}
\right]^{3(1-\beta)\over 2(1-3\beta)}\left[A_{{}_V} \eta^2+B_{{}_V}
\eta+C_{{}_V} \right]^{-{{1+3\beta}\over{2(1-3\beta)}}}~~,
\end{equation}
and
\begin{equation}
\label{h15}
 H_1 = {1 \over 3} {{a^3}^{\prime} \over a^3} = {a'_{1} \over a_1} =
 - {1 \over (1+3\beta)}
{ {(1 - 3 \beta) m_\beta \eta + \eta_0 } \over
( A_{{}_V} \eta^2 + B_{{}_V} \eta + C_{{}_V} )} ~~~. \end{equation}
The scale factors are
\begin{equation}
\label{a15}
a_1=\left[{-2(1+3\beta)}\over{[ 2+(1- \beta)(1+3\beta)\omega]}
{m_\beta}\right]^{1\over 2(1-3\beta)}
\left[ A_{{}_V}\eta^2+B_{{}_V}\eta+C_{{}_V} \right]^{1\over2( 1-3\beta)} ~~~,
\end{equation}
and
\begin{equation}
\label{a25}
a_2=a_1\exp\left[{-2h_2 \over \sqrt{\Delta}}
{\rm arctanh} \left[ {-2(1-3\beta)[(1-3\beta)m_\beta \eta + \eta_0] \over
(1+3\beta) \sqrt{\Delta}}\right]\right]~, \Delta > 0
\end{equation}
or
\begin{equation}
\label{a2n5}
a_2=a_1\exp\left[{ 2h_2 \over \sqrt{-\Delta}}
{\rm arctan} \left[ {-2(1-3\beta)[(1-3\beta)m_\beta \eta + \eta_0] \over
(1+3\beta) \sqrt{-\Delta}}\right]\right]~, \Delta < 0
\end{equation}
with
\begin{equation}
\label{a2a3}
a_2 a_3=a_1^2~~~. \end{equation}
These solutions are new.
However, for $\Delta=0$,
\begin{equation}
\label{aequal}
a_3=a_2=a_1 ~~~~~, \end{equation}
is obtained.
This last solution is clearly seen to be the one previously obtained for
the isotropic, FRW model, with an open space $(k=-1)$. Again, the $\Delta=0$
is obtained only if $h_1=h_2=h_3=0$.

Independent of the value $\Delta$ might have, $ h_1 + h_2 + h_3=0 $ is
always true.  In the Type V models with $ \Delta \not = 0$ one must have
that $h_2=-h_3$ with $h_1 = 0$.  For the latter case, truly
anisotropic solutions are obtained with
\begin{equation}
\label{dis5}
\Delta={B_{{}_V}^2-4A_{{}_V}C_{{}_V}} = {{-8 (1-3\beta)^2} \over
{18 \beta+(1+3\beta)^2 \omega}}{h_2^2}  \,\ .
\end{equation}

$C_{{}_V}$, being proportional to the sum of the squares of
the constants $h_2$ and $h_3$, carries the information concerning the nature
of the anisotropic character of this Bianchi-Type model.

Since equation (\ref{sig}) holds for all the Type V models, the
$\Delta \not= 0$ solutions could have had asymptotic behaviors to call them
"nearly isotropic in appearance'' if $\sigma \to 0$, when $\eta \to \infty$. In
this regard one finds that: for the $\Delta < 0$ solution $a_2$ tends to
$a_1 exp({\pi h_2} /{\sqrt{-\Delta}})$ and $a_3$ tends to
$a_1exp({- \pi h_2}/{\sqrt{-\Delta}})$ so, when $ \eta$ reaches the value
${(100 \sqrt{-\Delta} - B_{{}_V}})/2A_{{}_V}$ these two scale factors differ
from each other by one percent, and this solution is then ``ninety nine percent
near" the $ \Delta=0$ solution, corresponding to the  FRW cosmology for
$k = -1$ (see equations (\ref{phisol}) and (\ref{asol})).
On the other hand, for the $\Delta>0$
model the scale factors can never approach to a same, single, function of
$\eta$. The reason for this is that $\eta$ is bounded, see equation
(\ref{a25}).
Nevertheless, it is significant that this last can be an inflationary
solution.

\vskip .4cm
\noindent {\bf Bianchi type IX}
\par
 $\psi$ given by equation (\ref{psi}) is, likewise, a solution
for this model with the
$H_i$'s given by equation (\ref{hi}). However, in this case the $h_i$'s cannot
be constants. Instead, the $h_i's = h_i(\eta)'s$ are now new, and unknown,
functions of $\eta$. \par
With equation (\ref{psi}) substituted into
\begin{equation}
\label{curv5}
[H_i\psi]^{\prime}=[1+(1-\beta)\omega]m_\beta +a^{6 \beta} \psi
\,\, {}^*R_{{}_{i\,IX}} ~~, \quad {\rm i=1,2,3} \end{equation}
one must solve for
\begin{equation}
\label{hi5}
 h_i^{\prime}~=~a^{6 \beta} \psi {}^*R_{{}_{i\,IX}}~+~{{2A_{{}_{IX}}
-[2(2-3 \beta)+3(1- \beta)^2 \omega] m_{{}_ \beta}} \over{3(1- \beta)}}~~.
\quad {\rm i=1,2,3} \end{equation}
The sum of the above three equations,
\begin{equation}
a^{6\beta}\psi\,{}^*R_{{}_{i\,IX}}~=~{{2A_{{}_{IX}} -
[2(2-3\beta)+3(1-\beta)^2 \omega]m_\beta} \over(1-\beta)}
\nonumber \end{equation}
is, given explicitly in terms of $a_1, a_2$, and $a_3$,
\begin{equation}
\label{rel5}
{a_1^4+a_2^4+a_3^4-2(a_1^2a_2^2+a_1^2a_3^2+a_2^2a_3^2)\over 2a^{6(1-\beta)}}
={[2(2-3\beta)+3(1-\beta)^2\omega]m_\beta-2A_{{}_{IX}}\over{(1-\beta)
\psi}}~~,
\end{equation}
from which any chosen scale factor can be solved as function of the
other two remaining ones.

On the other hand, equation (\ref{con2}) gives
\begin{equation}
\label{hiquad}
h^2_1+h^2_2+h^2_3 \equiv~{ \rm K^2} = -{\omega^3\over 2(1-\beta)^2}
\left[ P\eta^2+Q\eta+S\right]~~,
\end{equation}
where the $P$, $Q$ and $S$ constants, given in terms of $A_{{}_{IX}}$,
$B_{{}_{IX}}$ and $C_{{}_{IX}}$, stand for
\begin{equation}
\label{p}
P= XA_{{}_{IX}} - [4A_{{}_{IX}} - Y](1-3\beta)^2m_\beta~~~,
\nonumber \end{equation}
\medskip
\begin{equation}
\label{q}
Q = XB_{{}_{IX}} - [4A_{{}_{IX}}\eta_0-2Y m_\beta\eta_0 +
2(1-3\beta)B_{{}_{IX}}](1-3\beta)m_\beta~~,
\end{equation}
\medskip
\noindent and
\begin{equation}
\label{s}
S= XC_{{}_{IX}} - [2\Delta+ 2(1-3\beta)m_\beta \eta_0 B_{{}_{IX}} -
Ym^2_\beta \eta^2_0]~~~;
\end{equation}
\medskip
\noindent where
\begin{equation}
\label{x}
X=3(1+3\beta)(1-\beta)^2\omega m_\beta + 6(1-\beta)m_\beta-
2(1+3\beta)A_{{}_{IX}} ~~, \end{equation}
\medskip
\noindent and
\begin{equation}
\label{y}
Y= 2(2-3\beta)+3(1-\beta)^2\omega~~~. \end{equation}
\par
The isotropic model solution that belongs to this Bianchi-Type model is
obtained when $\Delta = 0$, where one has that $h_1(\eta)=h_2(\eta) =
h_3(\eta) = 0$ (see section 2).

Under any circumstance the functions $h_i$'s, which must still obey the
condition $ h_1 + h_2 + h_3  = 0 $, determine the anisotropic character
of the solutions.
\par
The $h_{{}_i}$'s can be given as
\begin{equation}
\label{h19}
h_1 =-\left[{\kappa^2 +4\kappa +1\over3( \kappa^2 + \kappa + 1)}
\right]{\rm K} ~~~, \end{equation}

\begin{equation}
\label{h29}
h_2 = \left[{-\kappa^2 +2\kappa +2\over3( \kappa^2 + \kappa + 1)} \right]
{\rm K} ~~~, \end{equation}
and
\begin{equation}
\label{h39}
h_3 = \left[{2\kappa^2 +2\kappa -1\over3( \kappa^2 + \kappa + 1)} \right]
{\rm K} ~~~,
\end{equation}
where now $\kappa$ is another, new and yet unknown, function of $\eta$:
Unfortunatly, for $\Delta\neq 0$ we were not able to obtain
the explicit functional dependence of $ \kappa = \kappa(\eta)$.
Even so, an asymptotic isotropic behavior, similar to other models, for the
present solutions is also expected based on the strength of equations
(\ref{sig})
and (\ref{hiquad}).
\section{DISCUSSION AND CONCLUSIONS}

The JBD cosmological equations for perfect fluids with barotropic equations of
state is seen capable of beeing displayed, through the use of reduced
variables,
in a way which first permits one to obtain non trivial, significant
solutions with little effort and next, but more important, to express them
in terms of the single function $\psi\,=\, A \eta^2 + B \eta + C$.  The fact is
that the aforementioned solutions belong to a class which embraces
Bianchi--Type models some of which, in turn, comprise the FRW isotropy
groups.
Moreover, stated explicitly this class contains the general (analytic)
matter solutions for the Bianchi--Type I model as well as solutions
for the other two Bianchi--Types examined in this paper, which in turn
include special ones, a subclass, that tend aymptotically, as $ \eta \to
\infty $, to corresponding FRW solutions.
The reason for the existence of this set is that the functional form of the
product of the scalar field, $\phi$, times a power of the mean scale factor,
$a^3 = a_1a_2a_3$, as a function of the time parameter $\eta$ is a solution
to the equations used in this work independent of the metrics that give rise
to any possible present anisotropy for Bianchi--Types I, V and IX models,
and it has the FRW form.
In other words, for a perfect fluid with a barotropic equation of state, we
have shown that there exists a class of solutions for the Bianchi I,
V and IX types that contain their corresponding FRW models. The Type V
solutions are new, as well as those for the Type IX, but in the latter case
because of the complexity of the curvature terms it is only possible to give
the explicit form, in terms of $ \eta$, of the scale factors $a_i$'s
up to the single, unknown, function $\kappa = \kappa(\eta)$. Nevertheless
if $ \eta \to \infty$, an asymptotic isotropic behavior for the $\Delta \neq 0$
solution should be expected in view of equations (\ref{sig}) and
(\ref{hiquad}).
Moreover, there are also other solutions obtained from $\psi$, through equation
 (\ref{psi}), which describe other Bianchi--Type models, see
Chauvet and Guzm\'an \cite{ChGu86}.
All of the above are salient and remarkable properties of the class of
solutions that we have found. Even more, the $\psi$
solutions also include the asymptotic solutions, these are the Type I models
($\Delta =0, >0$ and $<0 $ ), for all Bianchi--Types near the initial
singularity, when the spatial curvature terms can be neglected. They are in
this sense --up to the matter terms-- comparable to the Kasner vacuum solution
of EGR.

At this point we want to mention that, in JBD, the separate cosmological
models are to be distinguished between themselves in a first instance,
through the different values that the constants $A$, $B$ and $C$
obtain. Also significant is the fact that $B$ and $C$ carry the physical
information on the nature of the presence, or even the absence, of the
anisotropy that any given models may have. Remind that these constants also
determine the value of the discriminant $\Delta$ in terms of the physical
parameters $\beta$ and $\omega$.
We stress the fact that solutions for the Bianchi models corresponding to
$\Delta = 0$ that possess an isotropy group, recover the FRW solutions: Type I
goes into the corresponding flat FRW one, Type V into the open FRW one and
the Type IX into the closed FRW one. Meanwhile, for the anisotropic solutions
with $\Delta \neq 0$ the models isotropize in the course of their time
evolution
by tending in the Type I case to its corresponding FRW model, while for Types
V and IX their evolution is toward the $\Delta = 0$ solution which is the only
one available to describe a non-flat FRW cosmology in the JBD context.

The present observational evidence points to a high degree of isotropy of the,
assumed relic, CMB radiation and, if so, it is a decisive argument in favour
of the, nowadays, large scale homogeneous and isotropic expansion of the
Universe. Then, if the initial stages of the expansion had a homogeneous but
anisotropic behavior one could follow, within the JDB, how an actual nearly
isotropic expansion can come about.
\bigskip

 {\bf Acknowledgment} This work was supported by CONACYT grant
400200-5-3672E.
\newpage
%%%%%%%%%%%%%%%%%%%%%%%%%%%%%%%%%%%%%%%%%%%%%%%%%%%%%%%%%%%%
%                      Bibliografy                        %
%%%%%%%%%%%%%%%%%%%%%%%%%%%%%%%%%%%%%%%%%%%%%%%%%%%%%%%%%%%%

%
\end{document}